\documentstyle[amscd,amssymb,12pt]{amsart}
\input xypic \xyoption{curve}
\input epsf

\def\hcorrection#1{\advance\hoffset by #1 }
\def\vcorrection#1{\advance\voffset by #1 }

\vcorrection{-.5in}
\hcorrection{-.8in}

\newcommand{\B}[1]{{\bold#1}} 
\newcommand{\C}[1]{{\mathcal#1}} 
\newcommand{\D}[1]{{\mathbb#1}} 

\begin{document}

\setcounter{tocdepth}{3} 

\pagestyle{plain}
\addtolength{\footskip}{.3in}

\title{On The Arrow of Time}
\author{Lucian M. Ionescu}
\address{Department of Mathematics, Illinois State University, IL 61790-4520
}
\email{lmiones@@ilstu.edu}

\maketitle

\begin{abstract}
The interface between classical physics and quantum physics is explained
from the point of view of Quantum Information Theory ({\em Feynman Processes}),
based on the {\em qubit model}.
The interpretation depends on a hefty sacrifice: 
the classical determinism or the arrow of time.
As a benefit, the wave-particle duality naturally emerges from the qubit model, 
as the root of creation and annihilation of possibilities (quantum logic).

A few key experiments are briefly reviewed from the above perspective:
quantum erasure, delayed-choice and wave-particle correlation.

The CPT-Theorem is interpreted in the framework of categories with duality
and a timeless interpretation of the Feynman Processes is proposed.

A connection between the fine-structure constant and algebraic number theory is
suggested.
\end{abstract}

\tableofcontents


\section{Introduction}\label{S:intro}
``It's been very difficult to make progress over the past 20 years,
because there hasn't been much new to say.'' says physicist Laura Mersini-Houghton,
co-organizer of the conference "The Arrow of Time" \cite{CAOT},
as reported in \cite{MSFT}.

If we recall that each unification of space, time and matter led to a new breakthrough,
e.g. the {\em Special Theory of Relativity} (\cite{Wiki}, a)) partially unifying space and time, 
and then {\em General Relativity} (\cite{Wiki}, b)) incorporating matter to space-time, 
we will recognize the importance of understanding the foundations of physics
at the {\em level of ideas}.
After re-evaluating the main concepts and fundamental principles,
but in accordance with, and as a result of the
feedback from the new mathematical-physics theories expressed at a {\em technical level},
we should be able to {\em say something new}, even if it is not the whole theory in final form.

Yet the crucial progress was achieved when a {\em unified description} was provided
for system observed {\em and} observer: quantum mechanics.
As a next step, 
its universal time was relativistically upgraded in Quantum Field Theory (QFT),
and even a space-time duality was mathematically incorporated into conformal field theory.

The interpretation of quantum mechanics did not keep up with the mathematical-physics
advancements (e.g. string theory is in need of an interpretation, the measuring ``paradox''
with its quantum-classical divide awaits for a resolution etc.).
The ``... tension between quantum mechanics and space-time theory'' \cite{AS}, p.119,
and many other conceptual difficulties,
are clear indications of ``the need for a deeper physics'' (loc. cit.).
Instead, considerable effort is consumed in trying to ``save'' the ``old physics'',
by incorporating quantum mechanics into a classical deterministic theory
\cite{NS-Aug407}, p.10.

Fortunately all the pieces of the puzzle are ready to be set in place for a 
``big picture'' resolving all the tension, 
provided we are willing to enlarge our 
``deterministic picture'', as suggested by the mathematics of QFT;
and the key concept at stake is (again) ``Time''.

In this article we will review briefly the quantum computing perspective on
space, time and matter, as different aspects of {\em Quantum Information Theory}: 
the theory of quantum data and programs \S\ref{S:STMI} 
(for additional details see \cite{UP,DWT1,DWT2}).

Next we will point to the wave-particle interpretation of the {\em qubit model}
of particle physics (Feynman Processes \cite{I-FL,I-FP}), 
with its precursor: spin networks and space-time foams \S\ref{S:PWD}.
An interpretation ``trick'' disposes of the ``collapse of the wave function''
as entanglement of external observables.

The other main idea in this article consists in taking the CPT-Theorem at face value:
{\em the time arrow is a natural choice of orientation of our quantum computation}.
It's precursor is Feynman's interpretation of anti-matter as matter traveling backwards in time,
as explained in \S\ref{S:CPT}.
It is not easy to get comfortable with this interpretation, 
so we ``practice'' in \S\ref{S:exp} reinterpreting some typical quantum experiments 
(double slit, quantum erasure, delayed-choice etc.).

In the concluding section \S\ref{S:conclusions}, 
we translate the consequences of our interpretation 
in a less technical language.
That ``Time'' is related to our perception process, 
and conscience could be responsible for the collapse of the wave function,
are not new ideas. 
Only now they have reached a reasonable maturity,
being ready to receive a satisfactory scientific formulation.
As a ``confirmation'' of the present author's views exposed in 
\cite{I:AOT}, an idea starts to spread \cite{NS:TU}: ``It is not reality that has a time flow,
but our very approximate knowledge of reality.
Time is the effect of our ignorance.''. 

\section{On Space, Time, Matter and Information}\label{S:STMI}
The benefits of adopting a computer science perspective on quantum physics
are multiple \cite{DWT1,DWT2}.
Furthermore, we advocate that it is captured by the ``grand unification'':
{\em Quantum Interactions are Quantum Communications}.
This prompts to the fact that space and time are mere coordinates
in a quantum computation \cite{UP,DWT1,BC-What-flows},
which could exist provided the quantum information flow is not ``turbulent'', 
as it will be explained bellow.

\subsection{Space and Time}
Mathematically, the above point of view amounts 
to model a quantum interaction as a representation of a {\em Feynman category},
called a {\em Feynman Process} \cite{I-FP,I-FL},
as for example in QFT by ``decorating'' graphs with propagators yielding 
probability amplitudes under specific Feynman rules, 
or String Theory which constructs the amplitudes from embeddings of Riemann surfaces
in an ambient space-time, as the constituents of the geometric Feynman category 
being represented.

It is crucial to understand that ``Space'' models the capability of parallel computation 
(tensor product $\otimes$ as ``horizontal composition''),
while ``Time'' models the capability of sequential computation 
(composition of morphisms as ``vertical composition'').
The ``rest'' of our image of space, which can be obtained as homogeneous space 
associated to the group of rotations $SO(3)$,
is encapsulated into the spin/qubit formalism of $SU(2)$.

Both space and time may be confined to a local existence, 
provided we disregard quantum corrections,
and which we view as ``turbulent flow'' of quantum information 
associated to loops (quantum information feedback leading to control and stability).

In a representation of tangles for example \cite{Turaev}, 
which will be explained in \S \ref{S:cat-model}, 
the input-output global computation can be factored in time-slices,
but the information flow can move back relative to a global vertical coordinate 
which plays the role of time.
The space-time structure is not preserved in such a quantum process,
where space (external degrees of freedom DOFs) appear in pairs via products and coproducts,
modeled usually by creation and annihilation operators.

\subsection{Consistent histories}
A more general quantum computation (QC) of Feynman type, over graphs for example, 
could be factored as a network of subgraphs which include loops, in such a way that
collapsing the quantum processes (viewing them as ``black boxes'',
i.e. quantum gates),
what remains is a representation of a distributive lattice conform to 
a classical computation satisfying our classical logic.
This ``picture'' is in the author's opinion the essence of the 
{\em consistent histories} interpretation of quantum mechanics
(see \cite{Histories} for an introduction).

Within the quantum computations performed by the black-boxes (``quantum chips'',
rather then just elementary quantum gates),
there is no admissible space-time coordinate system,
yet the qubit-lines should be conserved, the same way quark-lines are conserved
in elementary particle physics diagrams.

The lack of a time coordinate, even locally,
is reflected in the theory as the need to specify an order product expansion in QFT.
It is also supported by experiments which are a clear indication of the failure
of the classical model of a continuum space-time:
the defiance of light speed in photon tunneling \cite{NS-Aug1807}, p.10 
and \cite{SCIAM-Sept07}, p.50 
(which is in our interpretation a random walk in time - see \cite{DWT2}),
or the balanced homodyne detection \cite{QMC}, p.196 (noise interference
which in our interpretation is due to quantum space-time fluctuations),
entanglement and motion as a simultaneity issue \cite{NS-Aug407}, p.10:
`` ... a well-behaved time-ordering 't Hooft needs simply doesn't exist:
there is no causality at a deep level.''
\footnote{Causality is used here as of the ``light-cone type'', 
not in the sense of quantum correlations in a network
which {\em represents causality}, as in Feynman Processes \cite{I-FP}. }.

\subsection{Matter}
What flows, i.e. ``Matter'', 
is quantum information \cite{HDR-QDR}, with its elementary constituent,
the qubit, as an incarnation of the quantum non-exclusive dichotomy: ``Yes {\em and} No''.
Symmetry, responsible for mass, and its generator {\em entropy},
is interpreted as ``program'' acting on ``data'' \cite{DWT1,DWT2}.

A related quantum mechanics perspective on the 
information flow can be found in \cite{BC-What-flows}.

The role of the qubit as {\em the} elementary particle is proposed in \cite{DWT2}.
Here we will focus on a ``higher level'' interpretation of the qubit.

\section{Particle-wave duality and the qubit model}\label{S:PWD}
The wave-particle duality is essentially the ability of classical models
to explain some quantum phenomena, by switching between the {\em wave model}
and {\em particle model} as needed, 
but not within the same experiment!

The wave model is suited when {\em new DOFs are created} during the quantum experiment,
e.g. alternative paths, entangled spins etc., which subsequently entails 
{\em interference}.
We consider the splitting/creation of new DOFs (external or internal) as the
essence of the corresponding phenomenon ({\em delocalization}), 
while the fringes of interference,
or merely the oscillatory dependence of the probability of detection,
are consequences, well suited to be explained by a wave model,
e.g. Schrodinger wave mechanics, de Broglie pilot-wave etc. 
(coproduct/extension \cite{I-pQFT}).

The particle model captures {\em localization} of quantum events, as well as
their distribution as quanta, suited to be counted (Heisenberg matrix mechanics
as a dynamics of representations of quivers etc.).
We consider this merger/destruction/collapse of DOFs as the essence of the corresponding
phenomena, while the counting process is just a recording of this ``collapse''
of the external structure (product/quotient).

The creation and annihilation of DOFs is the trademark of QFT.
We interpret the wave-particle duality as resulting from a 
predominance of one aspect over the other in typical experiments.
There are though some experiments which involve both aspects
\cite{SKE,HC}.
The qubit model, where the unit of information 
$$SU(2)\cong\D{H}_1\ni Q=z_1|Yes>+z_2|No>, \quad z_1^2+z_2^2=1,$$
has both capabilities of splitting with interference, and localization with
counting (a number of qubits is defined),
and therefore exhibits both aspects, of wave and particle.

The {\em visibility} $V$ as a measure of interference capability (wave)
and the {\em knowledge} $K$ as a measure of (path) localization (particle)
satisfy $V^2+K^2\le 1$ \cite{SKE}, corresponding to the natural 
visualization of norm one quaternions $\D{H}_1\cong SU_2\cong S^3$ represented
as two disks: projection of upper and lower hemisphere \cite{DWT2}.

The qubit ($SU_2$) is essentially the ``double binary entity''
capable of splitting in ``space'' (parallel computing) and 
recombining in ``time'' (sequential computing) due to its structure:
$$\D{H}_1=\B{R}\otimes (\D{Z}_2\times \D{Z}_2)/\sim.$$
The resulting model generalizes Heisenberg's picture, i.e. matrices as 
representations of quivers, and its continuum model QFT,
to {\em Quantum Information Dynamics},
where the emphasis is not on the amplitude of probability (S-matrix etc.)
but on the actual {\em flow of energy-matter/quantum information}:
{\em Feynman Processes} \cite{I-FP,I-FL,DWT1,DWT2}.

The alternative incarnation is in the traditional Riemann surfaces
(Conformal Field Theory: CFT),
yet with the burden of continuity and ambient space-time 
which is a departure from the modern goal:
to build a discrete theory of finite type (yet not finite),
whether we believe the reality is (quantum) digital or not,
since we need to discretize our model to do the computations
and simulations on a computer anyway.

\section{CPT-Theorem taken seriously}\label{S:CPT}
The concept of determinism is deeply rooted in science, 
culminating with the classical mechanics point of view.
The determinism of Quantum Mechanics (QM), in terms of deterministic 
amplitudes of probabilities was never enough to satisfy the urge for 
the ``lost paradise'' of classical physics, leading to periodic attempts to
derive QM from a deterministic theory expressible in classical terms:
space-time, trajectory etc. \cite{DWT1,NS-Aug407}. 

We propose an indirect way to ``save'' the determinism, with a hefty price though:
sacrifice the {\em arrow of time} as an intrinsic concept.

Recall the CPT-Theorem of QFT: a simultaneous reverse of {\em parity, charge and time}
leaves the laws of physics unchanged.
Let us take this theorem seriously, as if we really mean it ...

\subsection{Co-determinism}
As pointed out by Feynman, anti-matter (charge conjugation) can be interpreted
as matter moving backwards in time.
We extend this idea by claiming that,
in some processes, {\em quantum deterministic matter evolving 
forward relative to our time-arrow is classical deterministic anti-matter
moving backwards in time}.
A bit more technically, the Feynman graph framework of QFT
leads to composition of trees and 
cotrees.
This leads to the Hopf algebra framework of quantization and related aspects
\cite{DWT2}.

Intuitively, we, as cognitive processes, {\em define} an arrow of time:
``{\em Cogito ergo sum}''.
In our ``quiet paradise'' most other processes are 
``emitters'' rather than ``absorbers'' of quantum information,
except when deciding to set-up quantum experiments,
exhibiting {\em creation of internal structure};
in this case we model quantum deterministically 
processes which are deterministic relative to the opposite time-orientation:
our present meets their future before acknowledging their causes.
The resulting model is QM (QFT) together with the {\em collapse of the wave function}.
We in fact model possible outcomes {\em backwards relative to their causality order}
and when we meet the ``real cause'' by measurement, we have to adjust our model.

The above interpretation is not just another 
interpretation of the {\em same mathematical-physics}
generally accepted.
It is conceptually compatible with (and even demanded by) 
the interpretation of elementary particle processes
as quantum information flow 
(e.g. tangles, Feynman diagrams \cite{I-pQFT} etc. - see also \cite{BC-What-flows})
forward as well as backwards relative to a local time-arrow (externally defined).
{\em We} break the Input-Output symmetry (morphisms of a hermitian category \cite{Bakalov}) 
because we distinguish between our ``interior self'',
and the ``exterior world'' from which we acquire information in a
consistent manner, enough to determine the Input-Output orientation of our
mathematical-physics models.

\subsection{Matter-Antimatter}
As a surprising ``side-effect'', this also explains the ``matter-antimatter ``paradox'':
Why do we see only ``matter''? where is the ``anti-matter'', 
if the ``Big-bang'' was symmetrical?

The ``answer'' (in the context of the above interpretation) is: it is ``here'', moving
backwards in time.
Then there should be ``almost'' the same amount of quantum information moving deterministic
along our knowledge journey through the universe as well as quantum information
moving co-deterministic, which we ``swim'' against in the ``ocean of information''
(a tensor category with duality and hermitian conjugation \S\ref{S:cat-model}, \cite{Bakalov}).
If we compare the mass, as a quantity of quantum information or the equivalent amount of energy,
with the usual communication streams (in the order of terabytes probably),
we see that the universe is ``almost symmetrical'' relative to the matter-antimatter inversion.

\subsection{Collapse of the wave function}
The co-deterministic picture just decrees that the collapse of the wave function is legitim,
when the dynamics is dualised and modeled using coproducts and superpositions.
There is no absolute determinism; the classical determinism (``particles/ products/ functions'')
together with the quantum determinism (``waves/ coproducts/ trees'')
can be restated as a generalized determinism, relative to a ``local time-arrow''. 
Reality altogether is deterministic, yet we (or our models), as only a part of it,
cannot know everything to predict ``everything'', and what we perceive as
the future will always surprise us;
that's what the free will is for!

\section{The Quantum-Classical Interface: theory and examples}\label{S:exp}
In this section we ``test'' our interpretation on some characteristic quantum processes.

\subsection{Quantum erasure}
The quantum erasure experiment can be reproduced at home \cite{SCIAM-QE} 
(see also \cite{KE,TBST}),
or at least the double slit quantum interference \cite{VI-exp}. %
Its interpretation consists in manipulating the proportion of the ``wave model'' and
``particle model'' within the same experiment, in what appears to be a
temporal succession relative to observer's global time and arrow.

The first part of the experiment consists in a classical double slit experiment,
and we will talk about electrons being sent to the double slit screen and 
then scattering photons from a 2nd source, to distinguish easily between the various
qubits involved.

If the double-slit experiment is concluded 
without further interaction with the ambient/prepared environment,
it yields the well-known fringes of interference.
 
In {\em this case} the {\em interpretation} ``tilts'' towards a wave behavior: 
the electron ``takes'' both paths and corresponding to the differences in length, 
``part A'' of the electron,
which ``went'' through one opening interferes with ``part B'' of the electron,
which ``went'' through the other slit (one can in principle deem the source 
of electrons to produce an ``electron at a time'').

Now comes the observation (measurement/interaction) by ``shining'' a 2nd light source
({\em not coherent} with the 1st source) at the double slit panel.
When an {\em interaction occurs} between a ``part'' of an electron and a photon (not a ``part''!?),
the ``which-way information'' {\em becomes available}.

In the cited references \cite{SCIAM-QE} the authors make the usual assumption:
observing {\em the} electron at one slit implies it went {\em only} through that slit.
``If we measure which of the two ways the particle went,'' (loc. cit.)
``we will only ever measure that it went {\em exactly} one of those ways.''.

Their statement amounts to a {\em reduction} of the wave-particle {\em model} of the electron 
to a particle model once the measurement (observation of the incoming electron) was made,
justified by the fact that the interference capability of the electron is ``gone''.
Here it is only at the level of the interpretation/modeling,
that the ``collapse'' from a wave to a particle model occurs;
in {\em fact} we don't know if the electron did not go through the other
slit too, sneaking unnoticed (not measured) to the final measuring site,
as it will be proposed shortly.

Reiterating, they assume that ``seeing'' {\em the} photon at the A-slit 
{\em excludes} an interaction at the B-slit (classical {\em XOR}).
This {\em is a localization} made at the level of the interpretation, i.e.
converting the model into a particle model.

Now, why admitting that the electron can split like a wave when interference is
recorded at the target, but deny this possibility (wave behavior) to the {\em second particle},
the photon, when it comes from the 2nd source?
In other words, why modeling the second photon as a particle only,
and not as a wave-particle?
$$
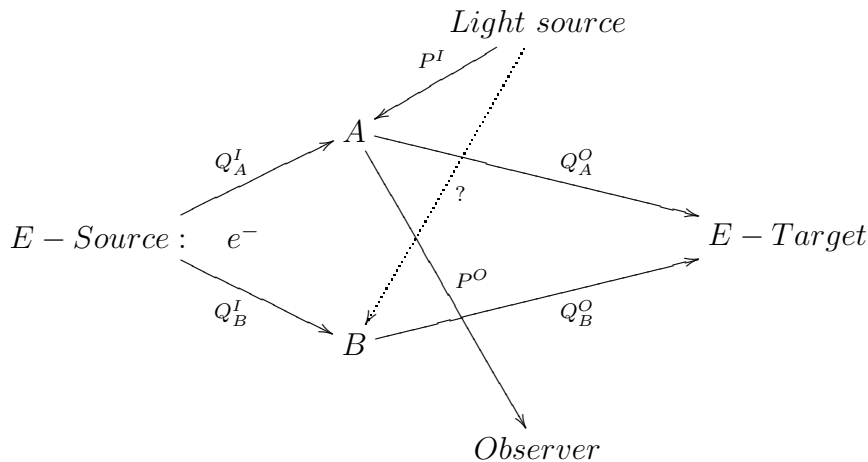
\diagram
       & & Light\ source \dlto_{P^I} \ar@{.>}[dddl]^{?} &\\
       & A \ar[drr]^{Q^O_A} \ar[dddr]^{P^O} & \\
E-Source: \quad e^- \urto^{Q^I_A} \drto_{Q^I_B}& & & E-Target\\
       & B \ar[urr]_{Q^O_B} & & & \\
       & & Observer &
\enddiagram$$

\subsubsection{The quantum deterministic interpretation}
We propose an alternative {\em quantum deterministic} interpretation
which keeps the wave-particle duality,
using the qubit model, explaining in a different way the lack of interference
(``decoherence'' in some sense), but in this way maintaining the capability for 
interference to explain the restoration of interference by quantum erasure.

Specifically, the double slit acts as a {\em quantum beam splitter},
which produces a pair of {\em entangled qubits},
with a non-localized electron particle-wave character.
When they are measured at the final target $T$, they interfere producing the
fringes of interference (a quantitative investigation will appear elsewhere).
When the 2nd source produces a photon (one qubit for now) not coherent with the electron,
an interaction with the qubit $A$ represents a measurement which reduces
the entangled state, implying that the other qubit $B$ has ``opposite amplitude
of probability''.
In essence, the entanglement of internal DOFs, like spin, is extended
to {\em external DOFs} (paths), 
with the usual flavor of ``spooky action at a distance''.

At this point we mention the alternative to be explored: the interaction changes
randomly the phase of $Q_A$, since the two sources are 
{\em not synchronized},
and therefore it still does interferes with $Q_B$,
but randomly, so the coherent interference pattern is lost.
The scattered photon is entangled with qubit $Q_A$, so the ``scrambled''
phase can be restored sending the information carried by the photon to the 
measurement site, to restore the interference pattern.

Now taking seriously the {\em qubit model}, we {\em split the photon} (``the particle'')
into two qubits $P_A$ and $P_B$ as being also a ``wave-particle'',
corresponding to the two slits {\em viewed from the other 
side}.

The electron-photon interaction at the two slits subsystem results  
in the {\em entanglement} of external and internal information,
i.e. the which-way and the internal phase information,
of the two entangled qubits of the electron with the two qubits of the photon
(quiver isomorphic to the homotopy class of a QFT Feynman diagram \cite{I-rem}). 

In this way we avoid using the classical logic underlying statements like 
``a particle have traveled one way (X)OR the other''
when explaining the experiment.

In other experiments the polarization is entangled with the which-way information.
In fact, as mentioned before, splitting a qubit amounts to entangling the
which-way information with itself.

Entangling the external/internal DOFs and then ``{\em erasing}'' 
the which-way information is implemented
at the level of qubits by the rotations in the external basis or the internal $\D{C}$-space basis,
corresponding to the complex structures of the qubits \cite{DWT2}.

Regarding the erasure part, 
it seems to be a process of eavesdropping using light
on the quantum communication using electrons:
$$
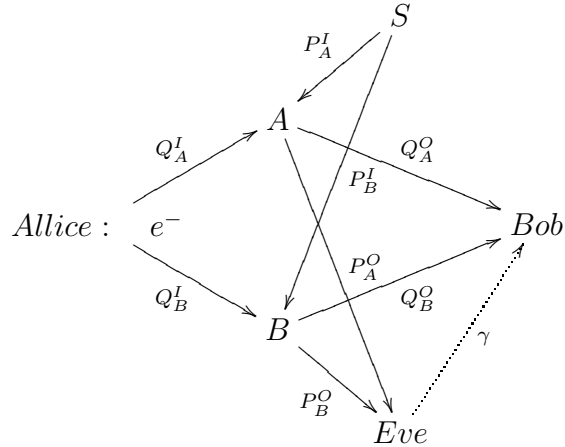
\diagram
       & & S \dlto_{P^I_A} \ar[dddl]^{P^I_B} &\\
       & A \ar[drr]^{Q^O_A} \ar[dddr]^{P^O_A} & \\
Allice: \quad e^- \urto^{Q^I_A} \drto_{Q^I_B}& & & Bob\\
       & B \ar[urr]_{Q^O_B} \drto_{P^O_B}& & & \\
       & & Eve \ar@{.>}[uur]_{\gamma}&
\enddiagram$$
together with a teleportation achieving the erasure.

The eavesdropping reads part of the quantum information, which due to the
no-cloning theorem, does not reach Bob anymore.

With no lens to focalize the two photon-qubits (\cite{SCIAM-QE}), 
their interference destroys part
of the classical information needed to be sent to Bob to complete the teleportation
of the ``missing quantum information''.
If the lens is present, then the two qubits interfere constructively
and sending the classical information completes the teleportation,
allowing Bob to recover the coherent interference pattern
(see {\em teleportation} and {\em Bell measurements} \cite{Tele}).

\subsubsection{The co-deterministic interpretation}
A two slit experiment can be equally viewed backwards relative to our time-arrow:
an anti-electron from what we call ``the target''
is split, taking the two paths, interfering at what we call ``the source''.
Mathematically, this corresponds to modeling the experiment in the opposite category,
or in the context of hermitian categories, by taking the adjoint,
as it will be explained in \S\ref{S:cat-model}.

At this point the reader may object: ``what if we consider a point on the target
where there is destructive interference; there is no ``photon'' there!? 

In the qubit model,
two qubits with ``opposite'' amplitudes of Yes and No, 
may ``collide'' and cancel the {\em measured response} on the measuring apparatus,
i.e. the amplitude of probability for Yes,
without annihilating each other, and conserving the quantum information.

This is similar to the use of complex numbers in order to represent alternating electric current,
resulting in a conservation law via unitarity.

The No amplitude of the qubit is not detectable, and may be thought of as ``dark matter'',
not necessarily having to do with the correspondent concept in cosmology.
The author believes that {\em that} has to do with symmetry (mass and information: another ``story'').

Then an ``anti''-electron, viewed as a qubit, white-matter or dark-matter,
is emitted by the target, splits into two qubits when passing through the screen,
and is absorbed by the source.

Now if we ``look'' and see the electron at slit $A$, on the target's side, 
then it must have been white-matter, and we have perturbed one qubit,
while the other passes through the slit B, 
and no longer interferes at the source, 
appearing relative to our time as {\em two} distinct electrons.

In conclusion, the interpretation uses the capability of the qubit model to allow the 
``wave function'' to collapse under the projection from qubit to amplitude,
corresponding to a measurement acting as a filter ``Is it true that ...'' and accepting
the Yes component, while conserving the quantum information.
In a destructive interference, the amplitude of the ``favorable case'' (Yes alternative,
relative to a measurement basis) will ``vanish'' from the observer's point of view (apparatus recording), 
being ``rotated'' on the ``dark side'' (No), as if it would be ``dark matter''.

Reversing the time orientation does not bring anything new in this case, 
since the geometry of the two slit experiment is invariant, 
unless we view it as a loop!

\subsubsection{The ``timeless'' model}
The idea is to reverse time only on one path, in order to {\em track the ``particle''} \S\ref{S:CSIS},
not the intersections of its ``world line'' with our time-advancing space-like slices
of reality.

Let's describe the experiment starting with our time-arrow in mind.
So, the electron emerges from the source and ``takes {\em one} path'' towards the screen,
passes through whole $A$, bounces on the source $S$ as an ``anti''-electron (time reversal),
may go through the second whole $B$ (one alternative) and reach the source as an ``anti-electron''
moving backwards in time, which usually interpret as a {\em second} electron (qubit) ``taking''
the second path {\em simultaneously}!
$$
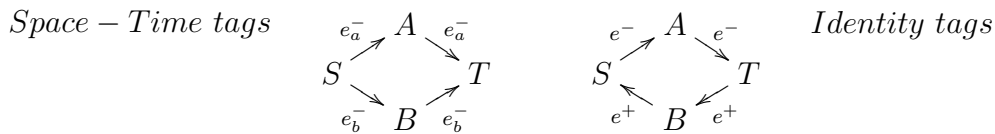
\xymatrix @C=12pt @R=2pt {
Space-Time\ tags &  & A \drto^{e^-_a} &  &  & & A \drto^{e^-} & & Identity\ tags\\
& S \urto^{e^-_a} \drto_{e^-_b} & & T & & S \urto^{e^-} & & T \dlto^{e^+} \\
&  & B \urto_{e^-_b} & &  & & B \ulto^{e^+} &
}$$

In other words, we have a {\em time-like loop of {\bf one} qubit}, 
if we accept that changing charge due to time reversal does 
not change the {\em identity} of the quantum subsystem (More on this issue later: \S\ref{S:CSIS}). 

So, in some sense we have restored ``determinism'', or rather the particle picture,
since the wave-behavior will appear as a resonance or self-interference phenomenon
on such closed paths.
This leads to the description of quantum phenomena as a Hodge-de Rham theory on graphs
and their complexes ({\em Quantum Dot Resolution} \cite{DWT1,DWT2}).
At the level of only one Feynman graph ({\em interaction mode}),
the physics is similar to what happens with electric circuits, as modeled
by Kirchoff's Laws and solved using Maxwell's methods:
currents may flow through the loops {\em even when there are no external sources}
\cite{Sternberg}.

Summing up, the {\em Qubit Model} promotes Feynman alternative paths to foreground:
the ``bunch'' of paths ``is the wave'', and each of them
carries quantum information, e.g. the qubit thought of as ``particle''.
The ``upgrade'' of the formalism is modest: from paths carrying amplitudes to 
paths carrying qubits (spinors), 
but the benefits are considerable, including a smooth and intuitive interface 
with our everyday language, the interpretation of the theory.

If we wish to make it look like a ``particle theory'',
we can disregard the space-time coordinate system and use 
a {\em Quantum Identity System} instead \S\ref{S:CSIS}.
Is then the theory deterministic?
Yes and No! 
The theory has a statistics component due to quantum fluctuations ``off shell''
which on average give the ``on shell'' physics laws.

This is similar with what happens in any {\em communication
with noise}, as described by {\em Shannon's Theory} \cite{Ion-TI},
due to the language issue: we are in fact trying to understand reality
by asking questions in a permanently adapting language.

\subsubsection{Towards Infotronics}
The doubling of QFT from an ``S-matrix theory'' with coefficients in $\D{C}$,
to a {\em Quantum Information Theory}
as a dynamics of the qubit flow with non-commutative coefficients in $\D{H}=SU_2$,
perhaps could also accommodate the dichotomy white-matter/dark-matter,
like extending from the holomorphic to include the anti-holomorphic part.

Returning to \cite{SCIAM-QE}, the authors acknowledge that 
``Curiously, in the field of quantum computation, quantum physicists seem
less hesitant to claim that the qubits are in a state of being ``0'' and ``1''
at the same time. 
Such statements are completely equivalent to saying that
a particle passed through one slit and the other slit simultaneously.''; quite right!
We are claiming that this is the case, i.e. results in a better model,
{\em independent of weather they have been measured or not}!

In other words, at the modeling level, the possible transitions (paths) and 
interactions are the ``reality'' (space-time-matter), as far as we can ``understand'' from
the corresponding observation.

\subsection{Delayed-choice experiment and wave-particle correlation}
Similar interpretation assumptions occur in the {\em delayed-choice experiments}
\cite{RS,Wheeler-GE,NS-Aug407}.

First proposed by Wheeler as a ``gedanken experiment'' \cite{Wheeler-GE},
it was actually performed in various experimental arrangements
(A. Zeilinger \cite{SCIAM-Aug07}, A. Suarez \cite{NS-Aug407}, P. Aspect and others).

Here again we advocate for the qubit model, quantum circuits \cite{DWT2},
input-output ports, not just modulation of amplitudes of probability \cite{Wheeler-GE},
and {\em against} the assumption present in the interpretation
``Thus {\em the} photon has traveled only one route'' (loc. cit. p.967).

Wave-particle correlation experiments \cite{SKE} rule out the naive wave-particle duality
as a possibility to switch between the wave model and the particle model as needed,
as if quantum mechanics is a ``split extension'' of wave mechanics and particle mechanics.

A quantitative explanation of visibility (interference) and which-way-information (localization)
in terms of correlated/anti-correlated qubits will be given elsewhere.

\subsection{Physics: modeling ``change''} 
The root of the apparent conflict between classical and quantum physics is 
the difference in scope: classical physics is mainly concerned with {\em motion},
i.e. change in the relative capabilities of interactions between ``objects'',
with its basic concept of {\em position}, while quantum physics and modern science
are concerned with {\em processes}, i.e. changes in the structure of systems
due to interactions, to the point where the ``identity'' of the subsystem is lost,
and a ``new subsystem is born'' as a result of an exchange during the interaction.

Focusing on the identity of elements is characteristic of set theory, while change in
the structure, e.g. considering {\em extensions} is characteristic of category theory,
focusing on relations (zoom in on a system, insertion and collapsing subsystems/subgraphs,
products {\em and} coproducts etc.).


At the level of mathematics, the change is not so drastic, 
since the main ideas are carried over, as it will be explained next.
Roughly speaking, the geometric dualism of the Hilbert space framework for 
Quantum mechanics with its conserved number of particles (identity),
is ``upgraded'' to the framework of categories with duality,
where the pairing (``gluing'' an object and its dual) plays the role of the
scalar product in a pre-Hilbert space, while the {\em copairing},
also called a unit, plays to role of a Hilbert basis, 
i.e. a complete orthonormal system.
In addition to the conceptual advantages explained above,
from the technical point of view this modern ``mathematical technology'' is 
considerable more versatile and computationally tractable, since it replaces
``hard-analysis'' (theory of unbounded operators etc.) with an algebraic 
framework of {\em finite type}.
For a concrete, but highly specialized example, see \cite{KS}.

Returning to the time-arrow issue,
a mathematical model for the CPT-transformation based on hermitian categories
will be explained in \S\ref{S:cat-model}, as an alternative of the global-classical
approach via the various components of the Lorenz group.
This later classical framework belongs to a ``pointwise physics''
based on the idea of a classical coordinate system.
The critique of the concept of coordinate system \S\ref{S:CSIS} 
will prepare the stage for the modern ``physics of processes'',
with its natural object-and-relations oriented language: category theory.

\section{From Hilbert Spaces to Hermitean Categories}\label{S:cat-model}
The traditional Hilbert space formalism quite adequate for Quantum Mechanics, 
with its conserved number of particles, is no longer sufficient as a language to 
express the qualitative features of Quantum Field Theory: 
the variable number of particles and change of types of particles (superselection rules and sectors),
in correlation with entanglement (non-locality).

The more {\em refined and rich language} is that of categories with duality.
The (new) paradigm consists in a geometric-topologic category of objects
like points or strings and corresponding  Feynman graphs and Riemann surfaces
as transitions, representing the ``external structure'' of the system under consideration,
which will be represented in an algebraic {\em coefficient category} of objects like spinors ($SU_2$-modules)
and transitions like unitary operators, having the role of ``internal structure''.

The ``two good examples making a theory'' are: electric circuits and Maxwell (homological algebra)
methods \cite{Sternberg} and Conformal Field Theory (CFT), with its easier topological version,
Topological Quantum Field Theory.
To see how the non-linear geometric-topological structure reflects the linearized algebraic one,
see \cite{Ion-note-Frobenius}.

String Theory is the classical-modern hybrid, ``stuck'' in an ambient absolute space-time
of the Einstein type, no matter how many dimensions might have.
To ``free its power'', it needs to be formulated as a Feynman Process, conform with the
current program of the present author \cite{I-FP,I-FL}.

\subsection{Duality in Categories}\label{S:duality-cat}
The sketch of the basic framework follows. This section mainly provides the keywords
and links to a detailed technical account \cite{Bakalov,Turaev},
while providing an intuitive description and motivation for their use in the 
physics application.

A category with duality is a braided tensor category $(\C{C}, \otimes, \sigma)$
where the familiar ``free adjunction of systems'' (``counting'' the input subsystems/ingredients in a
quantum process/computation/communication) modeled as a tensor product $\otimes$ 
is supplemented by a {\em braiding} which ultimately encodes the statistics.
A breading is a morphism in the category, for example $\sigma_{X,Y}:X\otimes Y\to Y\otimes X$,
which satisfies the relations implying the Yang-Baxter equation essential in the quantization process.
A ``bosonic'' example is the mere transposition $\otimes(x\otimes y)=y\otimes x$,
while a ``fermionic'' example is the supersymmetric commutative case $\otimes(x\otimes y)=-y\otimes x$.

Now comes the data which encodes creation and annihilation of particles.
A {\em ribbon category} (\cite{Bakalov} p.29) is a braided tensor category $\C{C}$
as above, together with dual objects $^*V, V^*$ and {\em pairings} $e_V:V\otimes V^*\to k$,
and copairings $i_V:k\to V\otimes V^*$, satisfying some natural axioms; here $k$ is the ground field,
say $k=\D{C}$.

The pairing generalizes the scalar product in the pre-Hilbert space formalism, 
while the copairing replaces the analytic condition of completeness with an algebraic counterpart;
it plays the role of a {\em complete Hilbert basis}.
Pairings and copairings also implement the creation and annihilation processes (operators).

{\em Modular categories} exhibit a compactness character at the level of {\em state sums},
and can be obtained from classical categories like $SU_2-mod$ via quantization at a root of unity
for example.

A category with duality is a more general case then the ribbon (or rigid) category,
allowing for a left dual $^*V$ to be different from the right dual $V^*$.
Such a structure can come from the existence of a second involution on the category,
a {\em hermitian conjugation} (\cite{Turaev}, p.), generalizing the case of Hermitean scalar products
of the Hilbert space formalism. 
Such a structure at work is used in \cite{KS}, where one could just assume that the 
modular category coming from subfactors indeed has the required properties (axiomatic approach). 
This is what we call a {\em Hermitean category},
and additional details can be found at \cite{I-sem-TQFT}.

\subsection{Feynman rules the Turaev functor}\label{S:Turaev-functor}
Now let us take a simple example of a geometric-topologic category: 
the category of ribbon graphs (see \cite{Turaev} p.32).
They are ``fattened'' Feynman graphs, and correspond to Riemann surfaces,
at least topologically at this point (see \cite{Kon-lowdt}.
One would have to add a ``width'' of the ribbon to complete the correspondence,
via a metric with its corresponding conformal structure.
A combinatorial presentation of Riemann surfaces is presented in \cite{Stahl}
(see Ch. 4).
The Turaev's ribbon graph corresponds to the {\em rotation system} determining 
the embedding of the graph presentation onto the correspondingly glued Riemann surface
(see also \cite{BIZ}).

With such a ribbon category {\em and} a coefficient category like $\C{C}$,
there is a unique Feynman-like functor, satisfying a set of Feynman-like rules,
which associates to a ribbon
colored with simple objects and morphisms from the category (the types of particles)
a unitary morphisms representing the evolution of the process (time-flow from dynamics).
We call it the {\em Turaev functor} \cite{Turaev}, Theorem 2.5, p.39.

Really important are the ideas, at this stage, not the details.
The classical paradigm of Field Theory, classical or quantum, based on the Lagrangian
formalism, even process oriented as in the Feynman path Integral formalism,
needs to be enriched by ``removing'' the presence of an ambient space-time
with a fixed topology, never mind the geometric dimensions.
This is a development of the language used to describe particle physics,
implementing the new ``discovered'' features of the fundamental forces
using a wider range of techniques, not ``confined'' to the gauge Theory paradigm.

This will be detailed further in \S\ref{S:elem-part}.
Before that, we need to critically reexamine the concept of a coordinate system.

\subsection{From Coordinate Systems to Identity Systems}\label{S:CSIS}
{\em Coordinate Systems} (CS) focus on how the position and orientation of the system
studied changes relative to the observer. 
They are useful to track momentum (position change) and angular momentum (orientation).
But, although these are conserved, they are rather a ``currency'' of interaction potential 
particles exchange, not the intrinsic characteristics of the particles, which are modeled
as internal properties: quantum numbers like lepton and barion number, spin etc.
The concept of CS, observer dependent but otherwise absolute and {\em particle independent}
leads to the concept of space-time via the principal of general covariance.
It leads to a total decoupling between observer and system observed: 
space-time is absolute and independent of the systems observed, as a homogeneous space
of a certain group of symmetries (Galileo or Poincare), and {\em The} system moves through it
(e.g. changes its position relative to a representative observer, i.e. CS,
while its irreducible components do not change (number of particles) and can be counted
or measured (measure does not change) etc.

The quantum world {\em is} change: ``Panta rei.''.
Particle physics adopted a system of tracking particles first, Feynman diagrams, 
and their positions next: Feynman rules.

At first, for Quantum Electrodynamics, it entailed only two types of particles 
which could {\em not} change one, electron as a fermion, into the other,
the photon as a boson (no supersymmetry).
Therefore the ``quantum identity'' (QI) of the particle was identical with the world line,
except for electric charge when matter annihilated antimatter (electron-positron annihilation).
Feynman's intuition recognized that more important is the QI, not the world line,
``interpreting'' an antiparticle as a particle moving back in time.
But he had enough trouble convincing the world that the path integral formalism, 
although without a mathematical solid foundation
\footnote{Saunders MacLane was not Feynman's classmate, as Grossman was for Einstein \cite{EG}!},
was the right way to model quantum {\em processes}.

After the modern advances in particle physics it is clear that one should track
the quantum identity of particles, including electric charge.
Without going into more details regarding how a new ``non-standard model'' must be designed,
reconsidering the ``fundamental interactions'' and corresponding charges together
with hindsight in view of the moderne knowledge acquired,
we propose the concept of {\em Quantum Identity System} which maybe intuitively thought
as world lines of constant quantum identity numbers.
Then there is the observer's external and classical CS in which the process is described:
the space-like and time-like of the measured events is specified in this CS,
corresponding to the above interpretation of {\em consistent histories}.
Inside a quantum process there is no time-order.
Only the QIS numbers are defined, and to relate with the phenomenology the usual
summation over paths (transitions) and time-orderings is performed.

\subsection{Information flow and lepton-quark numbers}\label{S:elem-part}
A representation of a Feynman diagram (graph with identity labels and Feynman rules) 
should be rather understood as a link diagram,
but {\em not} in $R^{3,1}$ space-time, but rather directly into say an $SU(3)-mod$
category, and the ``projection on the external ``space-time'' (causal structure in fact),
must be viewed as shadow the same way a knot diagram in the plane needs the additional
under/over-crossing information to characterize the ``original''.
In particle physics such diagrams represent the so called {\em quark lines}, but
they should include the lepton lines too.

A suggestive example is again the Turaev functor representing ribbons drawn in a ``space-time plane'',
where ``space'' really means the ``direction'' of the tensor product (monoidal structure),
and ``time'' is composition of morphisms.
The two operations $\otimes$ and $\circ$ can be unified within the framework of two categories
as {\em horizontal} and {\em vertical composition}.

Now the quark lines together with the phenomenon of confinement should and can be 
implemented as built in the model, {\em not} as a consequence of the strong ``force'',
as it was suggested in \cite{DWT2}, in connection with Wick rotation and a total
unification between space and time (see ``Tripling time dimensions'').

Then the lepton and quark (barion) number conservation will follow 
mathematically from the factorization properties of the category.
An example at the level of quantum mechanics can be found in \cite{BC-KQM}.

Such a factorization of a quantum process (quantum computation) into parallel 
and sequential subprocesses determines a class of {\em consistent histories}.

The projection onto an external ``time-coordinate'' shares the same properties 
of a Morse function on a compact manifold for example, 
defining an index and leading to conservation laws (lepton and quark numbers).
The change of flavor due to the weak ``force'' should (and can) be implemented 
using a homological algebra machinery in the above framework.
For example, in the ribbon graphs model ribbons can be thought of as mesons,
with one border having an opposite time orientation and representing an anti-quark
etc. (see \S\ref{S:FD}).

\subsection{The CPT Theorem and $\D{C}^*$-categories}
In QFT {\em charge reversal} $C$, {\em parity reversal} $P$ and {\em time reversal} $T$ 
are discrete symmetries \cite{Kaku}, p.115.
They {\em appear unrelated} because they correspond to the connected components 
of the Lorentz group \cite{W}, p.10.

In the more flexible framework of hermitian categories, 
$T$ is modeled by duality $V\mapsto V^*$, which reverses the morphisms,
which swaps the Input and the Output, therefore reversing the time arrow
by definition, conform to its present interpretation.
Charge conjugation $C$ is modeled by conjugation $V\mapsto \bar{B}$.
If the ground field is the field of complex numbers $k=\D{C}$, 
such a hermitian category will be called a $\D{C}^*-category$ 
(or unitary if of ``finite type'': {\em modular} \cite{Turaev}, p.113).

The terminology suggests the fact that results from $\D{C}^*-algebras$,
e.g. the theory of spectral triples of A. Connes and the theory of 
von Newman algebra ($W^*-algebras$), can be imported (``translated'') in the
more intuitive geometric-physics framework of hermitian categories
used as underlying Feynman categories 
(A word of caution: implementing multi-scale analysis,
i.e. resolutions, requires a differential: \cite{I-FP,I-FL}).

What is then parity reversal? Ribbon categories have an already made tool
to implement $P$: the ribbon twist $\theta_V$ \cite{Turaev}, p.20.
As emphasized in \cite{W}, p.8, 
``... that a physical system possesses space-inversion symmetry only makes sense
when one has specified what space inversion is supposed to do to the
observables of the system''.
If we think of the GNS-construction (\cite{Wiki}, c)), this corresponds to an operation
on the objects, representing states.
The CPT Theorem should be a constraint on the three involutions $*, ^{\_}, \theta$,
to the effect that $\theta_V:V\to V\cong V^{**}$ is a natural isomorphism in the 
$C^*$-category.
Note also that in the usual approach to QFT, where Fourier Transform $f(q)\mapsto \hat{f}(p)$
defines the duality between position representation and momentum-representation,
such a twist {\em is} parity reversal: 
$$\hat{\hat{f}}(q)=f(-q)=P(f)(q), \quad (FT=\sqrt{P}).$$
Again we emphasize that modeling the conceptual CPT-symmetries in a more ``varied''
way then just following the traditional group of transformations paradigm
provides additional qualitative features: it unifies them by allowing a ``smooth''
transition between time/charge/parity reversals (twisting and curling the ribbon),
which corresponds to the inexistent homotopy at the level of the Lorentz group.
More importantly, these ``topological moves'' 
together with a representation of the ribbon category using $SL_2(C)$,
which corresponds to the {\em reduced} Lorentz group \cite{W}, p.12, 
corresponds to the full Lorentz group.

There is still an additional ``bonus''. 
A classical way to unify (relate) 
the two connected components of the {\em proper} Lorentz group which includes
time reversal (but not parity reversal), is to consider the 
complex Lorentz group \cite{W}, p.13, and leading to the interesting twistor theory. 
This involves a {\em Wick Rotation} (WR) (a ``casual'' translation of QFT into an
Euclidean statistical quantum mechanics),
which applied twice yields the time reversal:
$$WR=\sqrt{T}.$$
to really unify space and time, one needs to allow $T$ and $P$ to be ``homotopical''.
This is achieved at the level of the categorical model via $C^*$-categories:
$$Duality\ in\ C^*-categories <=> Fourier\ Transform \oplus Wick\ Rotation.$$
These connections clearly require a more careful consideration and further research.

\section{Conclusions and Further Developments}\label{S:CFD}
In our opinion a comprehensive resolution of the conceptual tension in todays physics 
needs to address {\em all} the delicate issues, not just to focus on ``this or that aspect''.
Therefore mixing mathematical-physics and ``metaphysics'' can hardly be avoided.

\subsection{Conclusions}\label{S:conclusions}
The mathematics and physics have developed well enough for us to be able to give a solid interpretation
of quantum physics, i.e. relating our high-level language used to describe the quantum reality,
with the precise quantum mathematical-physics interface expressed in category theory language,
without our mind going `` ... ``down the drain'', 
into a blind alley from which nobody has yet escaped.'' \cite{F1}. 

The first main idea in this article is to focus on the {\em qubit model} ($SU_2$-representations
without ambient space-time) as providing the primary object of study,
the qubit flow (``non-commutative space-time''),
instead of the S-matrix approach focused on amplitudes of probabilities.
This determines us to claim ``reality'' for paths as well, and more generally,
for transitions as ``routes'' taken by the quantum information when it splits 
into ``packets'', as the ``best strategy'' when a quantum communication is in progress in the
{\em World's Quantum Web}.

The {\em qubit model implements the wave-particle duality}:
the wave characteristics come from a qubit representing space-like information (parallel processing)
and the particle characteristics (counting/localized detection etc.) comes
from the qubit's phase representing a local (Lie algebra/group: $TS^1=\B{R}$) 
``entropy arrow'' (input-output orientation).
It is an alternative to the CFT approach using Riemann surfaces,
delivering in this way the {\em s/t-duality} (at the mathematics level,
related to a generalized Wick rotation \cite{DWT2}),
a.k.a. wave-particle duality (at the physics level).
The creation of DOFs corresponds to creation of ``space'' and momentum, 
as a parallel quantum computing resource,
while destruction of DOFs (space) corresponds to creation of the dual resource, ``time''
and energy. 

The quantum erasure part will be explained quantitatively using the qubit model
elsewhere, in the spirit of {\em Infotronics}: the quantum engineering of 
quantum computations on quantum circuits/networks \cite{DWT2}.

The qubit model also pushes the quantum-classical divide to the ``objects'' 
(representing a collapse of DOFs; 
a clear-cut at the level of the model, but remember the resolution aspects when interpreting it) 
which model the sources and targets of the quantum transitions (the morphisms).
``All qubits arrive'' at the targets and interfere yielding a ``click''/count etc.,
the discreteness and randomness of which is an indication of the non-existence of a 
local time, not even discrete.

The second main idea consists in pushing further 
the (by now) orthodox Feynman's interpretation of antiparticles as particles traveling back in time,
taking the CPT-Theorem at face value:
the ``not classically deterministic matter'' is anti-mater traveling backwards relative
to our time-arrow determined by the Input-Output orientation of {\em our} quantum computation.
At the level of the mathematical model, this amounts to considering the opposite (hermitian) category
when interpreting portions of the consistent histories computation,
with the amendment that morphisms should be enlarged as pairs of morphisms,
in order to model relations (``functions and ``cofunctions''; quasi-isomorphisms of PROPs).

Regarding the ``time-reversal'' interpretation, the experiments involving detection without
interference (tree-like quiver) are classically co-deterministic,
which appear relative to our time-arrow as quantum probabilistic (pure states plus collapse
of the wave function).
The experiments which involve ``circuits'', exhibiting wave and particle characteristics
(in need of a combined wave-particle model, like the qubit model)
are quantum deterministic processes, reversible, 
therefore also quantum co-deterministic.

The matter-antimatter ``paradox'' vanishes, as matter is also antimatter, 
and our breaking of the CPT symmetry at low energies yielding the ``time'' symmetry 
(in quantum processes where time exists, or at least
can be organized in consistent histories) only alters the proportion slightly due to 
a natural choice of a time-arrow.

The measurement ``paradox'' seems to equally vanish if we accept that there is no absolute
determinism, and the universe of quantum communications from which we are only {\em a part},
will always affect ``our present'' in a co-deterministic way.

The qubit model combined with the time-reversal idea seems to prompt for the 
dichotomy {\em white-matter/dark-matter}, consistent to the doubling process
involved in the transition from an ``S-matrix formalism'' of QFT to the 
{\em qubit flow} of {\em Quantum Information Theory} and {\em Infotronics}.

As a side-effect of the mathematics needed by physics, 
it seems that a more symmetric treatment of classical mathematics is needed:
functions and cofunctions, Hopf objects (for non-commutative geometry) instead of fields (commutative geometry,
in need for quantization) etc. (see \cite{DWT2} for additional details).
At the more abstract (axiomatic/algebraic) level of category theory,
categories with duality provide a rich framework for particle physics,
allowing to shift the emphasis from studying motion to understanding the change of structure 
(including phase transitions). The ``upgrade'' from Quantum mechanics to Quantum Field Theory at the physics
level corresponds to the upgrade from Hilbert spaces to Hermitean Categories at the mathematics level.

\vspace{.2in}
The increasing use of discrete models (lattice models, graphs etc.) provides a short-cut
from the classical science of Galileo-Newton-Einstein where limits were mandatory to smooth out
computations, to the modern revival of the conceptual science (``Panta Rei'', Zeno and discrete space and time etc.)
freed-up of the necessary computational burden by computers as a counterpart to 
experimental technology as an interface to reality (Greeks never had their Galileo):
$$\xymatrix @C=4pt @R=2pt {
``Reality'' & Experiment & Interface & Theory\\
Classical & Galilei & Numerical\ Methods & Newton-Leibnitz-Einstein\\
Quantum & HEP \& LEP & Computer\ Simulations & Quantum \ MP\&CS.  
}$$

\vspace{.1in}
In conclusion, symmetry prevails and reassuringly, 
there is enough determinism making it worth planning ahead,
with a (qu)bit of surprise to make use of our free will 
to chose {\em our} future depending on what {\em it} has in store for us.

\subsection{Further Developments}\label{S:FD}
At this stage we only wanted to present a diversity of mechanism which should be used
as part of a richer language for particle physics, appropriate to design a model
beyond the Standard Model (SM), breaking the tradition with the gauge theory paradigm.

Preliminary considerations were recorded in \cite{DWT1},
and comments regarding the implementation of the ideas were given in \cite{DWT2}.

But the real fundamental question is what mass is and why the particles have the relative masses,
charges etc., measured \cite{W}, p.3. 
The answer should clarify what gravity is, as part of the overall picture,
and produce the coupling constants as derived quantities.
To be more specific, we will conclude with a specific example.

In \cite{DWT2} the idea that elementary particles are {\em primitive elements}
of the fundamental Hopf algebra $\D{Z}$ (what else!\footnote{``God made the integers''.}),
the {\em primes}, emerged from general considerations (see last chapters).
Since we consider mass as a generator of symmetry, and entropy as a measure of 
the group of symmetries, one would try to explain the logarithm of the relative masses of
particles (quarks for instance). 
A table is provided in \cite{DWT2} p.91.

Now the prime elements are the Lie algebra elements of $\D{Z}$.
Let us add them in ``parallel'', i.e. $\frac1R=\frac1{R_1}+\frac1{R_2}$,
as possible paths for the transition of a qubit 
(splitting qubits in a discrete manner, without continuous amplitudes).
Computing the ``parallel volume'' (space/cross-section) as a {\em probability}, 
since sequential processing is ``time'':
$$\frac1{P}=\sum_{p\in g(\D{Z})} \frac1{p^2} = 1/2^2+1/3^2+1/5^2+...\approx 0.434707,$$
we find a value close to (\cite{Wiki} b)):
$$I=\frac2{- \ln \alpha}= 0.40684, \quad \alpha=\frac{e}{2hc\epsilon_0}\approx 1/137.035999.$$
Refining the idea, since the quark color is related to a cubic, or possibly a six-th
root of unity, one should rather consider Gaussian primes instead;
but in what lattice, corresponding to $\D{Z}[i]$ ($\Lambda_4$, see \cite{PC}, p.22),
or $\D{Z}[j], j^3=1$ (lattice $\Lambda_6$, 
as a ``unified'' Galois extension containing the 2 and 3
dimensional extensions\footnote{The ``GUT group'' $SU_1\times SU_2\times SU_3$.})?

Summing the inverse of the norms of the ordinary primes 
which are {\em not split} in $\D{Z}[i]$ 
(2 which is ramified and the other inert primes \cite{PC,primes}), yields:
$$\zeta_g(2)= \sum_{p\ prime \in \Lambda_4, \ not \ split} 1/N(p)= 
1/2^2+1/3^2+1/7^2+1/11^2 ...\approx 0.398434,$$
with a relative error $|I-\zeta_g(2)|/I$ about $1.98\%$.

Even with such a mismatch, the confluence of the theory with 
that of the Riemann zeta function (see also \cite{Kr}), Galois theory and Riemann surfaces, 
Calabi-Yau manifolds and complex quaternions (Wick rotation) etc.,
gives such a beautiful picture, 
that ``it must be true'':
$$``Ultimate\ Physics\ Theory'': Algebraic-Geometric-Numbar Theory!$$.
If not yet convinced, I will invite the reader to notice the coincidence 
between the number of prime elements on the (``convolution'') diagonal
(loc. cit. p.32: 1, 8, 10, 14 ...), 
and the fusion rules of $SU_3$: $3\times 3\times 3=1+2\cdot 8+10+ ...$
corresponding to the classification of particles under the ``eight-fold way''!

Returning to the categorical picture, 
the ``parallel addition'':
$$R=R_1\oplus R_2, \quad \frac1R=\frac1{R_1}+\frac1{R_2}$$
is a natural counterpart of the concept of {\em distance},
viewed as an additive measure of ``resistance'' to interactions: the further in space the objects are,
the weaker the interaction is (confinement is a totally different issue),
and suited to measure the ``vertical distance'' in a tensor category thought of as a 2-category
of coefficients.
At the level of the geometric category, say of Riemann surfaces, the two additions are
related by inversion (e.g. the Riemann sphere as a {\em bifield} \cite{DWT2}).

Then the inverse of the zeta function $\zeta_g(2)$ plays the role of a total ``vertical width'':
$$N(g)=\bigoplus_{p\in g_{\ne split}} N(p).$$
In the context of {\em modular categories} (unitary with finitely many irreducible objects),
this seems to correspond to the {\em rank of the category} \cite{Turaev}, p.76:
$$Vertical\ Pythagora\ Theorem: \C{D}^2=\sum dim(V_j)^2,$$
with its ``bilinear polarized'' version,
The Verlinde-Moore-Seiberg Formula,
relating the statistics (the braiding) and multiplicities of the tensor product (fusion rules)
\cite{Turaev}, p.106.
But this ... is another ``story'' \cite{Ion-OM}.



\begin{thebibliography}{33}
\bibitem{primes} Wolfram's Mathematica was used for the computation, 
with primes<100000.
\end{thebibliography}
\end{document}